\documentclass[11pt]{article}
\usepackage[margin=20mm,a4paper]{geometry}
\usepackage{graphicx} 
\usepackage[hidelinks]{hyperref}
\usepackage{tikz}
\usetikzlibrary{patterns}
\usepackage{tikz-network}
\usepackage{multirow}
\usepackage{subcaption}
\usepackage{cleveref}
\usepackage[braket]{qcircuit}
\usepackage[hide]{ed} 

\usepackage{authblk}
\usepackage{cite}

\title{Quantum-Assisted Graph Domination Games}
\author[1]{C Weeks}
\author[1]{P Strange}
\author[2]{P Drmota}
\author[1]{J Quintanilla\footnote{Email: j.quintanilla\@kent.ac.uk}}

\affil[1]{University of Kent, Physics of Quantum \& Materials Research Group, School of Engineering, Mathematics \& Physics, Ingram Building, Canterbruy, Kent, CT2 7NH, United Kingdom}
\affil[2]{Department of Physics, University of Oxford, Clarendon Laboratory, Parks Road, Oxford, OX1 3PU, United Kingdom}

\date{November 19, 2025}

\begin{document}

\maketitle

\begin{abstract}
We study quantum advantage in the 1-step graph domination game on cycle graphs numerically, analytically and through the use of Noisy intermediate scale quantum (NISQ) processors. We find explicit strategies that realise the recently found~\cite{PiotrBasePaper} upper bounds for small graphs and generalise them to larger cycles. We demonstrate that NISQ computers realise the predicted quantum advantages with high accuracy.
\end{abstract}

\section{Introduction}

Quantum computing is an important change in the way we approach computation. A quantum processing unit uses quantum bits, or qubits, enabling them to exploit superposition, interference, and entanglement to solve problems that are impractical for classical computers due to excessive computation times~\cite{QCIntro}. 
While the term ``quantum advantage'' usually refers to the time complexity of a problem (as in the algorithms due to Shor~\cite{ShorAlgorithm} for factorisation and Grover~\cite{GroverAlgoritm}  for searching unstructured databases), qubits can also offer other types of advantage. For instance, quantum entanglement (namely, the ability of qubits to become linked in ways that create complex relationships between their states, even when they are far apart \cite{vedral2006introduction}) can be used to transmit a message with a guarantee that it has not been detected (quantum cryptography \cite{Pirandola2020Dec}) - a feat that is simply not possible using classical means. We could call such advantages ``quantum operational advantages'' to distinguish them for those of the purely computational kind. The present paper is concerned with a form of operational advantage, namely the use of entanglement to improve the probability of success of coordinated moves by distant parties without exchange of signals~\cite{brukner2006entanglement, RendezvousBase,tucker_quantum-assisted_2024,PiotrBasePaper}. In particular, we study recently introduced~\cite{PiotrBasePaper} quantum-assisted, two-player graph domination games, both theoretically and through simulation using classical as well as quantum processors. 

In graph theory, the domination problem is a purely combinatorial and static problem~\cite{HaynesHHS1998}. A dominating set in a graph is a set of vertices such that every vertex in the graph is either in the set or adjacent to a vertex in the set. The classical graph domination problem asks for the smallest such set, or more generally, for structural and complexity properties of dominating sets. Graph domination has many applications, for instance to facility location and coverage problems~\cite{FacilityLocation,Haynes2023-it}. Since the early 2000s, however, there has also been interest in dynamic variants of graph domination involving mobile agents constrained to move along the edges of the graph~\cite{Klostermeyer2020Oct}. For example, the eternal domination problem, introduced in 2005~\cite{GoddardHedetniemi2005}, models guards that must continually relocate in response to attacks while ensuring that the occupied vertices remain a dominating set. More recently, in distributed computing and mobile robotics, researchers have considered algorithms in which robots collectively move so as to form a dominating set, often under additional algorithmic or memory constraints~\cite{chand2023run}. Unlike classical domination problems~\cite{HaynesHHS1998}, the challenge in these agent-based domination problems is not only combinatorial (whether a dominating set exists) but also algorithmic and probabilistic (whether a team of moving agents can find one efficiently).

A recent development is the introduction by Viola and Mironowicz of a graph-domination game where players have access to entangled quantum resources~\cite{PiotrBasePaper}. In the game they considered, two players are placed randomly on the nodes of a graph. They are tasked with coordinating their movements in order to dominate as much of the graph as possible. A node is dominated when there is a player on that node or on a node connected to it by an edge. The degree of success of a given strategy is the number of dominated nodes after the players have moved. Strategies can be characterised by their ``domination number'': the average number of dominated nodes after many runs of the game. It was shown that this increases when quantum resources are available. Importantly, the qubits can be entangled before the players learn which vertices have been assigned to them. Afterwards, no further interaction or communication is required. The obtained quantum advantage is therefore not purely formal: it represents a real advantage that could be gained in real-life situations where such games need to be played, without the need to change the nature of the game - only the resources available to each player locally need to be modified.

In Ref.~\cite{PiotrBasePaper} bounds on the degree of domination for 2-player games were obtained through semi-definite programming. Graphs with up to 13 vertices were considered. In the case of the pentagon (graph $C_5$) the probabilities of all possible moves were reported. Here we build on that work by deducing explicit protocols that realise those probabilities and, for larger graphs, the predicted bounds. We generalise our strategies to cycle graphs of arbitrary size and simulate the games on NISQ hardware, finding that the qubits in present-day quantum processors are of sufficient quality to realise most of the predicted quantum advantage.

\section{Theory}
We start by defining the rules of the game. The graphs the players will move on are unweighted and undirected. Each player will be allowed one move from their random starting positions. These initial positions are taken from a uniform probability distribution (in particular, it is possible that the players start on the same location, though even then they do not know the location of the other player). Waiting is not an allowable action. After all players have moved, their collective domination of the graph is checked. The nodes of the graph are labelled. The players have access to the label of the site they are on and can use that label to determine their actions. In other words, the players have a shared 'map' of the graph. Finally, the players do not have to follow the same rules when making their decisions (player-asymmetric game).\footnote{This implies no loss of generality as player-symmetric strategies are a subset of player-asymmetric ones.} \\

Quantum advantage emerges when the players share parts of an entangled quantum system~\cite{PiotrBasePaper}. Specifically, each player has a member of a pair of entangled, $n$-state qudits, where $n$ is the number of edges for each node of the graph. After a player measures their qudit they use the result to decide which of the $n$ edges to travel along. Before measurement, each player is allowed to perform an arbitrary rotation on their qudit which may depend on their identity and the site they are on. If the two qudits are entangled, this crucial step allows players to encode information about their location in the shared quantum state - the essential source of the quantum advantage. 

In this paper, we will consider cycle graphs only, so the players will share a pair of qubits ($n=2$). The players always start by placing their qubits in the joint Bell state  
\begin{equation}
    |\psi\rangle=\frac{1}{\sqrt{2}}\left(|00\rangle+|11\rangle\right)
    \label{eq:Bell}
\end{equation} 
They then get separated and each player carries out a local unitary operation on their individual qubit that depends only on the site that player has been assigned. For the graphs we consider, the unitary operations can be assumed to always take the form of rotations around the $y$ measurement axis. 

Let us first consider the 5-vertex cycle graph $C_5$. This is the simplest non-trivial graph for the domination game.

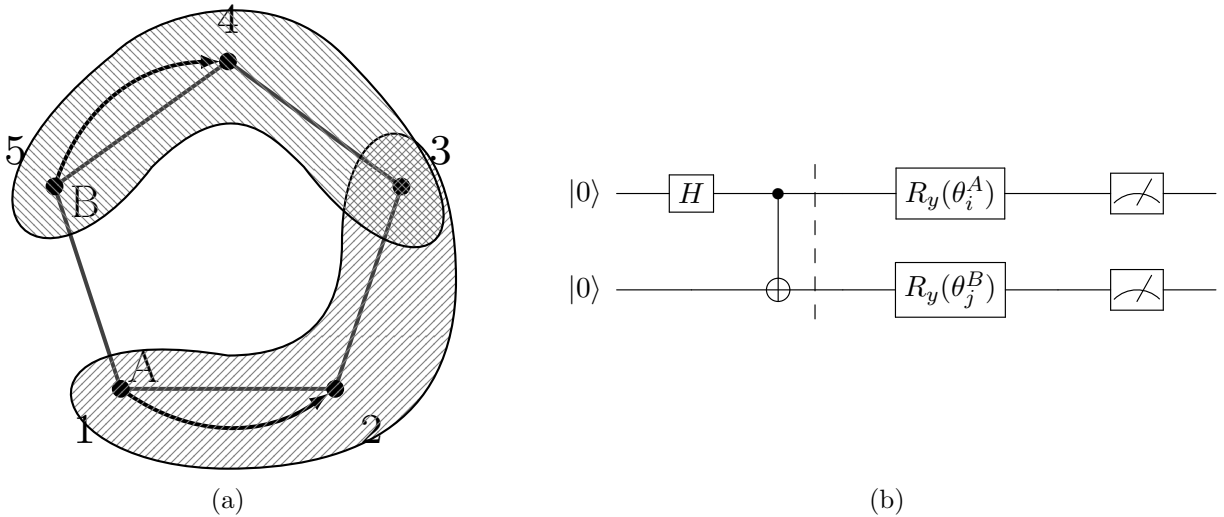
\begin{figure}[h!]
    \centering
    \begin{subfigure}[b]{0.5\textwidth}
        \centering
        \begin{tikzpicture}
            \Vertex[x=-1.410684, y=-1.94164,size=0.2,style={color=black},label=1,fontscale=2,position=234]{A}
            \Vertex[x=1.410684, y=-1.94164,size=0.2,style={color=black},label=2,fontscale=2,position=306]{B}
            \Vertex[x=2.282536, y=0.74164,size=0.2,style={color=black},label=3,fontscale=2,position=18]{C}
            \Vertex[x=0, y=2.4,size=0.2,style={color=black},label=4,fontscale=2,position=90]{D}
            \Vertex[x=-2.282536, y=0.74164,size=0.2,style={color=black},label=5,fontscale=2,position=162]{E}

            \node [font=\LARGE] at (-1.110684, -1.64164) {A};
            \node [font=\LARGE] at (-1.882536, 0.54164) {B};
            \Edge[bend=35,Direct=True,color=black](E)(D)
            \Edge[bend=-35,Direct=True,color=black](A)(B)
    
            \Edge(A)(B)
            \Edge(B)(C)
            \Edge(C)(D)
            \Edge(D)(E)
            \Edge(E)(A)

            \path[draw=black, thick, pattern=north east lines, pattern color=gray]
            (0,-3) .. controls (-2.5,-3) and (-3,-1) .. (0,-1.5) .. controls (1,-1.5) and (1.5,-1) .. (1.5,0) .. controls (1.5,2) and (3,2) .. (3,-0.5) .. controls (3,-2.5) and (1.5,-3) .. (0,-3) --cycle;
            \path[draw=black, thick, pattern=north west lines, pattern color=gray]
            (-1,1) .. controls (-2.5,-1) and (-4,0.5) .. (-1.5,2.5) .. controls (-1,3) and (0.5,3.5) .. (1.5,2.5) .. controls (4,0) and (2.5,-1) .. (1,1) .. controls (0.5,1.5) and (0,2) .. (-1,1) --cycle;
        \end{tikzpicture}
        \caption{\label{ClassPlay}}
    \end{subfigure}%
    ~ 
    \begin{subfigure}[b]{0.5\textwidth}
        \centering
        \Qcircuit @C=1.8em @R=1.5em {
                & \lstick{\ket{0}}  & \gate{H} &\ctrl{1} \barrier{1}&\qw &\gate{R_y(\theta^{A}_{i})} & \qw & \meter & \qw \\
                & \lstick{\ket{0}}  & \qw &\targ  &\qw &\gate{R_y(\theta^{B}_{j})} & \qw & \meter & \qw }
        \vspace{2cm}
        \subcaption{\label{2ChcoiceCirc}}
    \end{subfigure}
    \caption{Classical and quantum-assisted approaches to the graph domination game. (\ref{ClassPlay}) shows a play-through of the classical domination game. The players Alice (A) and Bob (B) are randomly placed (on sites 1 and 5, in this example). They then use a pre-agreed strategy to decide their moves. In our example, A moves from 1 to 2 and B moves from 5 to 4, which corresponds to the optimal classical strategy shown in Fig.~4 of Ref.~\cite{PiotrBasePaper}. They then check how much of the graph is dominated between them (shaded area). In this case, they dominate all 5 nodes of the graph. (\ref{2ChcoiceCirc}) shows the quantum circuit A and B can use to gain quantum advantage. Before A and B get separated (left of the dashed line) they perform a joint operation on their two qubits that places them in the entangled state of Eq.~(\ref{eq:Bell}). Afterwards the players get assigned their starting nodes and A(B) rotates her qubit around the $y$ axis by an angle $\theta_i^A$ ($\theta_j^B$) that depends on the index $i$ ($j$) of the site she is in. She then measures her qubit in the computational basis and uses the result to decide whether to move clockwise (1) or anti-clockwise (0).\label{fig:class_quant_game_illus}}
\end{figure}

The classical version of the game is illustrated in Fig.~\ref{ClassPlay}.%
The optimal classical strategy (shown in Fig.~4 of Ref.~\cite{PiotrBasePaper}) gives an average domination of 4.6 nodes.  The quantum approach for all cycles is illustrated in Fig.~\ref{2ChcoiceCirc}. For $C_5$, it gives a domination number of $\approx 4.76$~\cite{PiotrBasePaper}.

In order to find the explicit strategy that realises the optimal bounds found by Viola and Mirionowicz, we can employ a similar heuristics to that used for the rendezvous problem in Ref.~\cite{tucker_quantum-assisted_2024}. We start by creating the ``domination table'', Tab.~\ref{C5DomTable}, which shows the domination number depending on the sites Alice and Bob land on and on whether they move clockwise (1) or counter-clockwise (0).\footnote{A similar table for a larger graph ($C_{10}$) has been provided in the Appendix for illustration purposes.} A random coin-flip leads to an unweighted average of all the numbers on the table ($D=4.2$). The optimal classical strategy shown in Fig.~4 of Ref.~\cite{PiotrBasePaper} corresponds to picking a specific row (column) for each of Alice's (Bob's) possible starting sites. For instance, for the case where Alice starts on Site 1 we pick 0 (counter-clockwise motion) and for the case where Bob starts on Site 5 we pick 1 (clockwise motion). This is the situation illustrated in Fig.~\ref{ClassPlay} and leads to a higher average domination number $D=4.6$. Note, however, that this means Alice's and Bob's moves are uncorrelated: Bob will move clockwise whenever he starts the game at Site 5, independently of Alice's location. In the quantum game all combinations of starting sites and measurement outcomes are possible but they have different probabilities given by 
\begin{equation}
    \begin{pmatrix}
    P(00) & P(01)\\
    P(10) & P(11)
    \end{pmatrix}
    =\frac{1}{2}
    \begin{pmatrix}
    \cos^2{\frac{\theta^B_{j}-\theta^A_{i}}{2}} & \sin^2{\frac{\theta^B_{j}-\theta^A_{i}}{2}}\\
    \sin^2{\frac{\theta^B_{j}-`\theta^A_{i}}{2}} & \cos^2{\frac{\theta^B_{j}-\theta^A_{i}}{2}}
    \end{pmatrix}.
    \label{eq:Ps}
\end{equation}
Here, $P(\sigma_A\sigma_B)$ denotes the probability of Alice's reading being $\sigma_A$ and Bob's being $\sigma_B$ ($\sigma_A,\sigma_B=0,1$). It depends on the angles $\theta^A_i$ and $\theta^B_j$ which are functions of the indices $i,j$ of Alice's and Bob's respective landing sites. 
\begin{table}[]
\centering
\begin{tabular}{|cc|cc|cc|cc|cc|cc|}
\cline{3-12}
\multicolumn{2}{c|}{\multirow{2}{*}{}} & \multicolumn{2}{c|}{1} & \multicolumn{2}{c|}{2} & \multicolumn{2}{c|}{3} & \multicolumn{2}{c|}{4} & \multicolumn{2}{c|}{5} \\ \cline{3-12} 
\multicolumn{2}{c|}{} & \multicolumn{1}{c|}{0} & 1 & \multicolumn{1}{c|}{0} & 1 & \multicolumn{1}{c|}{0} & 1 & \multicolumn{1}{c|}{0} & 1 & \multicolumn{1}{c|}{0} & 1 \\ \hline
\multicolumn{1}{|c|}{\multirow{2}{*}{1}} & 0 & 3 & 5 & 4 & 4 & 3 & 5 & 5 & 4 & 4 & 5 \\ \cline{2-2}
\multicolumn{1}{|c|}{} & 1 & 5 & 3 & 4 & 5 & 5 & 4 & 3 & 5 & 4 & 4 \\ \hline
\multicolumn{1}{|c|}{\multirow{2}{*}{2}} & 0 & 4 & 4 & 3 & 5 & 4 & 5 & 4 & 5 & 3 & 5 \\ \cline{2-2}
\multicolumn{1}{|c|}{} & 1 & 4 & 5 & 5 & 3 & 4 & 4 & 5 & 3 & 5 & 4 \\ \hline
\multicolumn{1}{|c|}{\multirow{2}{*}{3}} & 0 & 3 & 5 & 4 & 4 & 3 & 5 & 5 & 4 & 4 & 5 \\ \cline{2-2}
\multicolumn{1}{|c|}{} & 1 & 5 & 4 & 5 & 4 & 5 & 3 & 4 & 4 & 5 & 3 \\ \hline
\multicolumn{1}{|c|}{\multirow{2}{*}{4}} & 0 & 5 & 3 & 4 & 5 & 5 & 4 & 3 & 5 & 4 & 4 \\ \cline{2-2}
\multicolumn{1}{|c|}{} & 1 & 4 & 5 & 5 & 3 & 4 & 4 & 5 & 3 & 5 & 4 \\ \hline
\multicolumn{1}{|c|}{\multirow{2}{*}{5}} & 0 & 4 & 4 & 3 & 5 & 4 & 5 & 4 & 5 & 3 & 5 \\ \cline{2-2}
\multicolumn{1}{|c|}{} & 1 & 5 & 4 & 5 & 4 & 5 & 3 & 4 & 4 & 5 & 3 \\ \hline
\end{tabular}
\caption{The domination table for the $C_5$ graph where players independently decide whether to move clockwise or anti-clockwise by flipping a coin (or examining a qubit). The number at the top of each column represents Alice's site, and the number at the front of each row represents Bob's site. 0 and 1 correspond to clockwise and anti-clockwise moves, respectively.}
\label{C5DomTable}
\end{table}

Any given quantum-assisted  graph-domination strategy is completely specified by the functions 
\begin{align}
    \theta^{A} & :i\mapsto\theta_{i}^{A}\\
    \theta^{B} & :j\mapsto\theta_{j}^{B}
\end{align}
These in turn determine the probabilities in Eq.~(\ref{eq:Ps}). Using these, we can straight-forwardly obtain the average domination number:
\begin{equation}
    \begin{split}
        D_5=\frac{1}{25}\biggl\{&3\sum_{i=1}^5\cos^2\left(\frac{\theta_i^B-\theta_i^A}{2}\right)+4\sum_{i=1}^5\left[\cos^2\left(\frac{\theta_i^B-\theta_{\left(i+1\right)}^A}{2}\right)+\cos^2\left(\frac{\theta_i^B-\theta_{\left(i-1\right)}^A}{2}\right)\right]\\
        &5\sum_{i=1}^5\left[\cos^2\left(\frac{\theta_i^B-\theta_{\left(i+2\right)}^A}{2}\right)+\cos^2\left(\frac{\theta_i^B-\theta_{\left(i-2\right)}^A}{2}\right)+\sin^2\left(\frac{\theta_i^A-\theta_i^B}{2}\right)\right]\\
        &\frac{9}{2}\sum_{i=1}^5\left[\cos^2\left(\frac{\theta_i^A-\theta_{\left(i+1\right)}^B}{2}\right)+\cos^2\left(\frac{\theta_i^A-\theta_{\left(i-1\right)}^B}{2}\right)\right]\\
        &\frac{7}{2}\sum_{i=1}^5\left[\cos^2\left(\frac{\theta_i^A-\theta_{\left(i+2\right)}^B}{2}\right)+\cos^2\left(\frac{\theta_i^A-\theta_{\left(i-2\right)}^B}{2}\right)\right]\biggr\}.
    \end{split}
    \label{eq:D5}
\end{equation}

Here, all site indices that are shifted (e.g. $i+1$) are all implicitly modulo 5, so as to account for the periodic nature of the cycle graph. 

We have optimised numerically Eq.~(\ref{eq:D5}) with respect to the 10 angles $\theta^A_1,\ldots\theta^A_5,\theta^B_1,\ldots,\theta^B_5$ using the Broyden–Fletcher–Goldfarb–Shanno algorithm~\cite{Scipy}. The results suggest that the optimal angles are given by
\begin{align}
    \theta^A_i&=\left(i-1\right)\theta_5
    \label{AliceAngEq}
    \\
    \theta^B_i&
    =\pi+\theta^A_i
    \label{BobAngEq}
\end{align}
where
\begin{equation}
    \theta_5=\frac{2\pi}{5}
    \label{AngSepEq}
\end{equation}
is a fixed angle increment. We have confirmed analytically that this is indeed a maximum, and it coincides numerically with the previously discovered~\cite{PiotrBasePaper} bound, $D\approx 4.67361.$ The difference of $\pi$ between the angles used by Bob and Alice ensures that the players spread out when they start together, which is obviously desirable for graph domination. Moreover, the probabilities obtained by substituting Eqs.~(\ref{AliceAngEq},\ref{BobAngEq}) into (\ref{eq:Ps}) are those given in Ref.~\cite{PiotrBasePaper} for this game.

\begin{figure}
    \begin{subfigure}[b]{0.5\textwidth}
        \includegraphics[width=\linewidth]{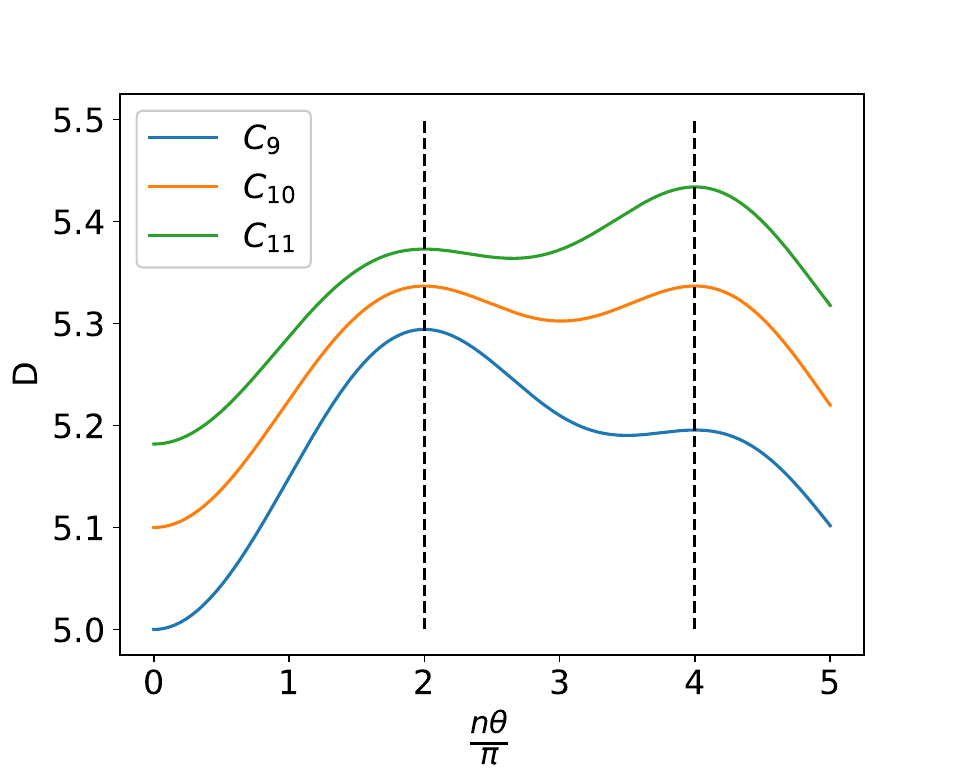}
        \caption{}
        \label{DEqGraph}
    \end{subfigure}
    \begin{subfigure}[b]{0.5\textwidth}
        \centering
        \includegraphics[width=\linewidth]{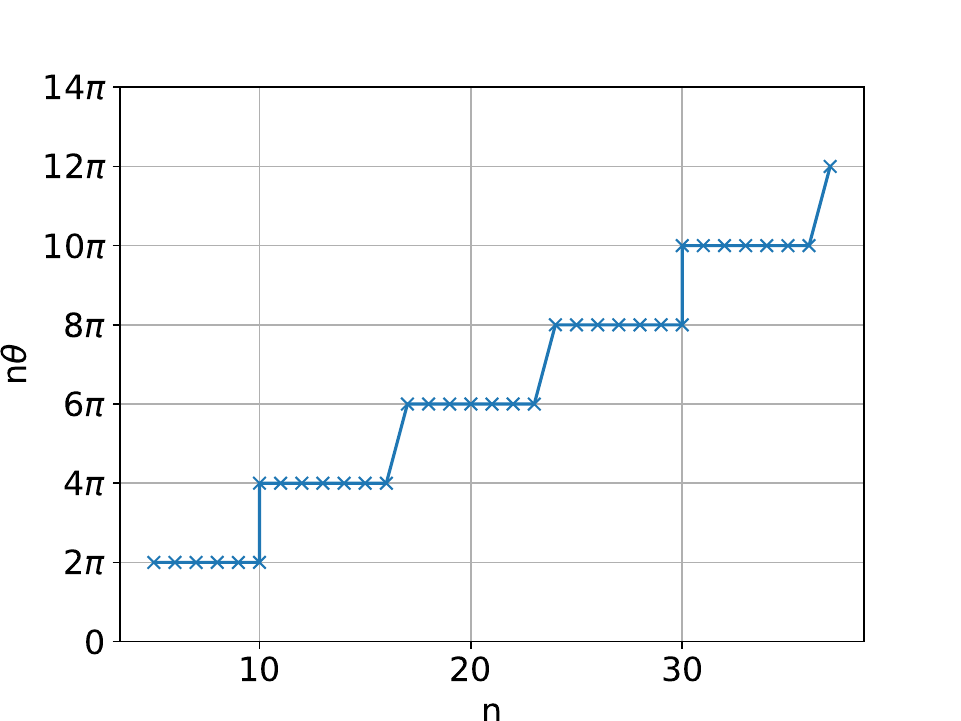}
        \vspace{0.5cm}
        \caption{}
        \label{optimal_theta}
    \end{subfigure}
    \caption{Optimization of the angle increment $\theta$ for maximising the domination number $D(n,\theta)$. (a) Domination number as a function of the angle for $n=9,10,11$, showing a shift in the optimal value of $\theta$ from the one given in Eq.~(\ref{AngSepEq}) ($n=9$) to the one given in Eq.~(\ref{AngSepEq2}) ($n=11$). (b) Optimal value of $\theta$ as a function of $n$. }
\end{figure}

We now generalise  (\ref{AliceAngEq},\ref{BobAngEq}) to 
\begin{align}
    \theta^A_i&=\left(i-1\right)\theta_n 
    \label{AliceAngEq_n}
    \\
    \theta^B_i&
    =\pi+\theta^A_i
    \label{BobAngEq_n}
\end{align}
and take it as an {\it ansatz} for graph domination games with $n>5$. This allows us to obtain a more general expression for the average domination number: 

\begin{eqnarray}
D & = & 6  + \frac{1}{n}\left(-8+\cos (\theta) - \frac{1}{2}\cos(2\theta) - \cos(3\theta) -\frac{1}{2} \cos (4\theta)\right) \cr
& & \cr
& + & \frac{1}{n^2} \left(-\cos (\theta) + \cos(2\theta) + 3\cos(3\theta) +2 \cos (4\theta)\right.\cr
& & \cr
& - & \left.2\cos ((n-4)\theta ) -3 \cos ((n-3)\theta) -\cos((n-2)\theta) +\cos((n-1)\theta)\right)
\label{DominanceFormula}
\end{eqnarray}
Here, $\theta \equiv \theta^{A/B}_{i+1}-\theta^{A/B}_i$ is the constant difference between the angle set by Alice/Bob when they land on site $i+1$ and at site $i$. The above expression can be written more compactly as 
\begin{equation}
    D_n(\theta)
    =
    \lambda_n + \sum_{l=1}^{n-1} \mu_l \cos(l \theta)
    \label{eq:Dn_compact}
\end{equation}
The constant $\lambda_n$ and the coefficients $\mu_n$ can be deduced from the longer form. For convenience, they are given in Table~\ref{CyclicCoeffTable} for the first 7 values of $n$. 
We have verified numerically that the above equation gives an optimal strategy for every value of $n$ for which the bounds are known~\cite{PiotrBasePaper} i.e. $5 \leq n \leq 13$. Interestingly, the angle increment is not always given by the straight-forward generalisation of (\ref{AngSepEq}),
\begin{equation}
    \theta_n=\frac{2\pi}{n}.
    \label{AngSepEq_10}
\end{equation}
That expression is valid for $n\leq 10$ only. For $n=11,12,13$ we find instead
\begin{equation}
    \theta_n = \frac{4\pi}{n}.
    \label{AngSepEq2}
\end{equation}
The transition is illustrated in Fig.~\ref{DEqGraph} which shows the optimization for $n=9,10,11$ (for $n=10$, both choices of the angle optimize the domination number).
Beyond $n=13$ the optimal domination numbers are not known~\cite{PiotrBasePaper}. We can, however, hypothesize that the optimal strategy is still consistent with Eqs. (\ref{AliceAngEq_n},\ref{BobAngEq_n}) and determine numerically the value of $\theta$ that optimizes (\ref{DominanceFormula}). Analysis of equation (\ref{DominanceFormula}) shows that the value of $n$ at which the steps in $n\theta$ occur is determined by the relative phase of the terms in $1/n$ and those in $1/n^2$. $n$ does not occur in the argument of the cosines in the terms in $1/n$ whereas it does occur in some of the terms of order $1/n^2$. Therefore the position of the peak in the $1/n$ terms is, to a good approximation, independent of $n$, while the peaks in the $1/n^2$ terms increase in number and move. Each time a minimum in the $1/n^2$ terms passes the peak in the $1/n$ terms a step in $n\theta$ occurs. This shows us that the steps will occur when $n$ changes by about 6.67 and that if a step occurs at a particular value of $n$ one will also occur at $n+20$. This is displayed in Fig.~\ref{optimal_theta} for cycles with $n\leq 37$.

\begin{figure}
    \centering
    \includegraphics[width=0.8\linewidth]{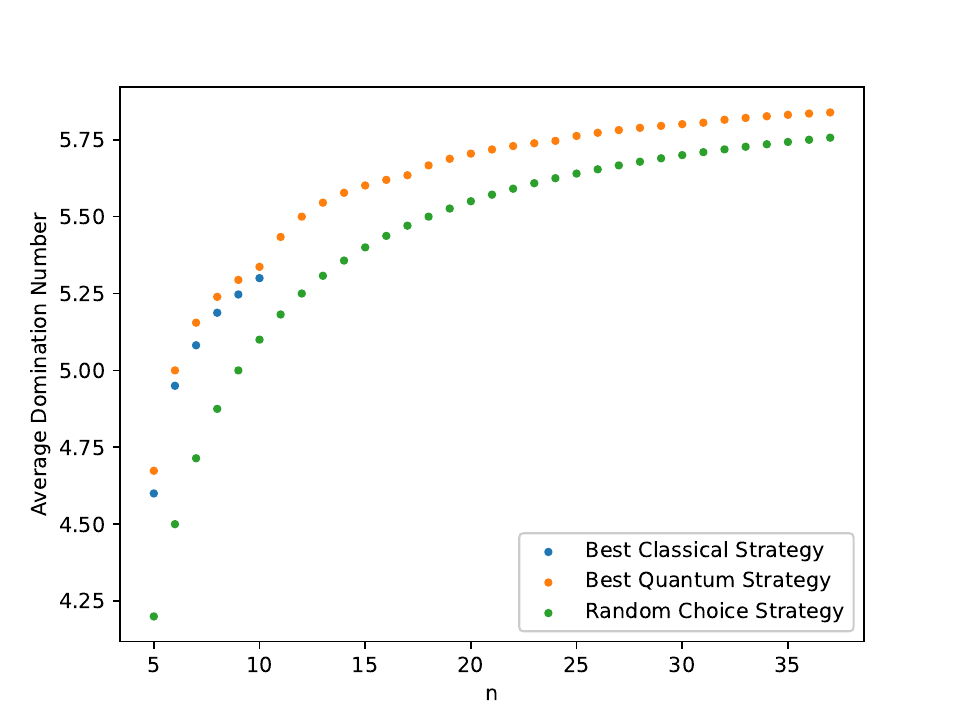}
    \caption{Theoretical prediction of the average domination number achieved by our optimised quantum strategy for cycle graphs compared to a coin-tossing strategy (random choice) and to the optimal classical strategy, when known. In each case, the domination number is given as a function of the number $n$ of vertices in the graph.}
    \label{fig:optimal_vs_n}
\end{figure}

The predicted performance for these optimized values of $\theta$ is shown in Fig.~\ref{fig:optimal_vs_n}. We cannot discard that better strategies might be available for  $n>13$.

\begin{table}[]
\centering
\begin{tabular}{|c|c|c|c|c|c|c|c|}
\hline
n & 5 & 6 & 7 & 8 & 9 & 10 & 11 \\ \hline
$\lambda_n$ & 85 & 138 & 203 & 280 & 369 & 470 & 583 \\ \hline
$\mu_1$ & 2 & 5 & 6 & 7 & 8 & 9 & 10 \\ \hline
$\mu_2$ & $-\frac{9}{2}$ & -4 & $-\frac{5}{2}$ & -3 & $-\frac{7}{2}$ & -4 & $-\frac{9}{2}$ \\ \hline
$\mu_3$ & -3 & -6 & -6 & -5 & -6 & -7 & -8 \\ \hline
$\mu_4$ & $\frac{1}{2}$ & -2 & $-\frac{9}{2}$ & -4 & $-\frac{5}{2}$ & -3 & $-\frac{7}{2}$ \\ \hline
$\mu_5$ &  & 1 & -1 & -3 & -2 & 0 & 0 \\ \hline
$\mu_6$ &  &  & 1 & -1 & -3 & -2 & 0 \\ \hline
$\mu_7$ &  &  &  & 1 & -1 & -3 & -2 \\ \hline
$\mu_8$ &  &  &  &  & 1 & -1 & -3 \\ \hline
$\mu_9$ &  &  &  &  &  & 1 & -1 \\ \hline
$\mu_{10}$ &  &  &  &  &  &  & 1 \\ \hline
\end{tabular}
\caption{Coefficients in Eq.~(\ref{eq:Dn_compact}) for cycle graphs with $5\le n \le 11$.}
\label{CyclicCoeffTable}
\end{table}

\section{Simulations}

In this section we will describe simulations of our graph-domination strategies using both classical and quantum processors. The former allow us to confirm convergence under ideal conditions.
The latter allow us verify that the predicted quantum advantages can be realised using current qubit technology. The {\it modus operandi} for these simulations is analogous to that used for quantum-assisted rendezvous in Ref.~\cite{tucker_quantum-assisted_2024}. 

\begin{figure}
    \centering
    \includegraphics[width=0.9\linewidth]{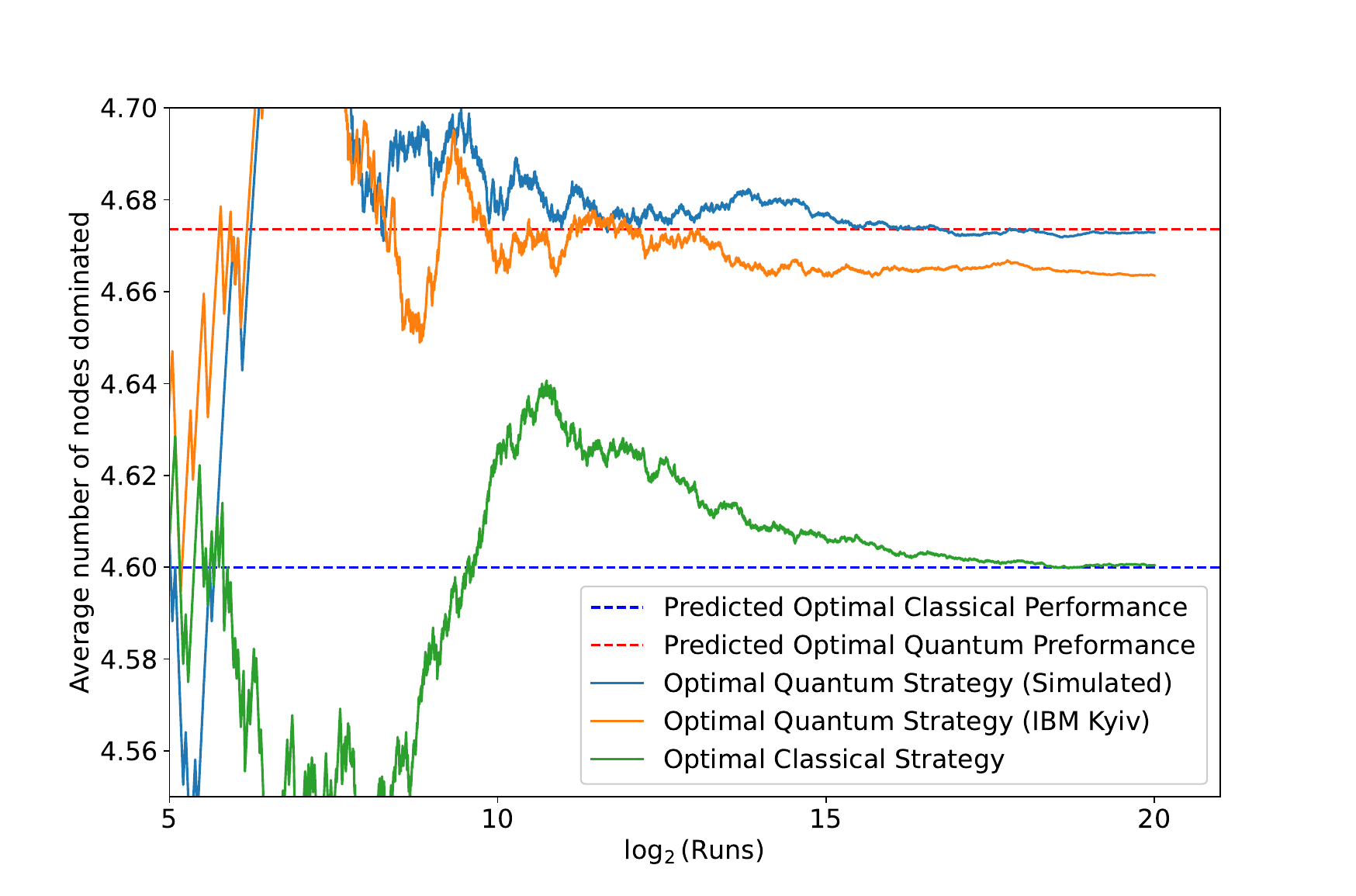}
    \caption{%
        Performance of optimal strategies for the graph domination game on a 5-site cycle. The dashed lines indicate the predicted averages over many runs for the optimal classical and quantum strategies, as indicated. The solid lines show: simulations of the optimal classical and quantum strategies using a classical computer; and a simulation of the optimal quantum strategy using the best found superconducting quantum processor, IBM Kyiv.
    }
    \label{C5Performance}
\end{figure}

\begin{figure}
    \centering
    \includegraphics[width=0.7\linewidth]{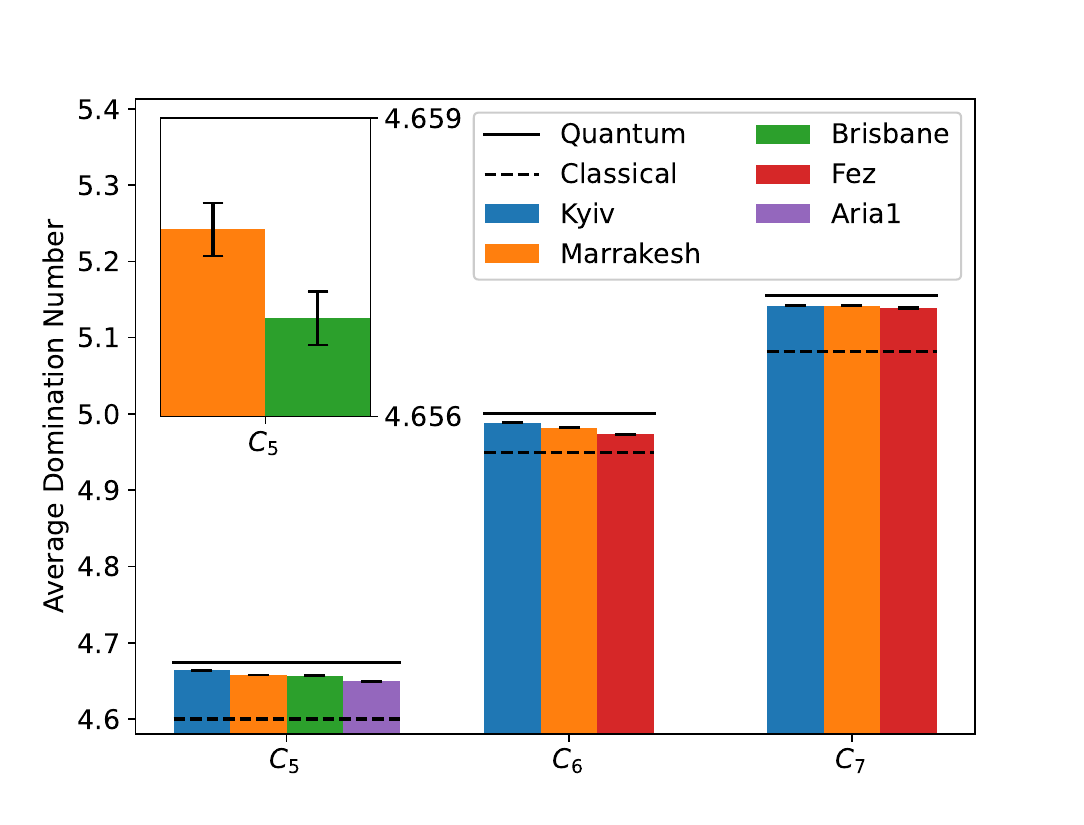}
    \caption{Performance of optimal strategies for the graph domination game on 5, 6 and 7 cycle graphs simulated on different quantum processors. The dashed line indicates the predicted average for the optimal classical strategy and the solid line indicates the optimal quantum strategy. The averaging is over $\sim 1000000$ runs in each case. The smaller lines represent error bars, estimated by the standard deviation of each averaged data set (a single line is visible due to the extremely small standard deviation, $< 10^{-3}$ in all cases; the inset shows the zoomed-in error bar for two of the quantum processors).}
    \label{CombinedPerformance}
\end{figure}

Our simulations of graph domination strategies on the simplest non-trivial cycle graph, $C_5$, are shown in Figure~\ref{C5Performance}. We observe clear convergence of the classical simulations towards the predicted average domination numbers, with the performance of the classical and quantum strategies clearly differentiated after averaging over about $\sim 1,024$ runs. 
Similar results were obtained for $C_6$ and $C_7$, albeit convergence was somewhat slower for larger graphs (see figures in the Appendix).  
For the quantum simulations we used four different superconducting quantum processors, namely \verb|ibm_kyiv|\cite{IBMQuantum_Kyiv}, \verb|ibm_marrakesh|\cite{IBMQuantum_Marrakesh}, \verb|ibm_brisbane|\cite{IBMQuantum_Brisbane} and \verb|ibm_fez|\cite{IBMQuantum_Fez}, as well as, in the case of $C_5$, one trapped-ions quantum processor, \verb|IONQ Aria1|\cite{IonQ_Aria1}. The converged results for the graphs $C_5, C_6,$ and $C_7$ after averaging over $\sim 1000000$ runs are shown in Fig.~\ref{CombinedPerformance}. \verb|ibm_marrakesh|,  \verb|ibm_brisbane|, and \verb|ibm_fez| show similar performance, with \verb|ibm_kyiv| somewhat better than the other three. \verb|IONQ Aria1|, used for $C_5$, shows performance below all three superconducting quantum processors. 
In all cases, there is a clear convergence towards a value that is much closer to the prediction for the optimal quantum strategy than the classical one. 

To quantify the observed quantum advantage $A$ we use the following formula:
\begin{equation}
    A=\frac{Q-C}{C-R}
    \label{AdvEq}
\end{equation}
Here $Q,C,$ and $R$ are the average domination numbers with the optimal quantum strategy, the optimal classical strategy, and random choice, respectively. The results are summarised in Table~\ref{tab:achieved_quantum_advantages}.
\begin{table}
\begin{centering}
\begin{tabular}{|c|c|c|c|}
\hline 
 & $C_{5}$ & $C_{6}$ & $C_{7}$\tabularnewline
\hline 
\hline 
\textbf{Predicted} & \textbf{18\%} & \textbf{11\%} & \textbf{20\%}\tabularnewline
\hline 
IBM Marrakesh & 14\% & 7\% & 16\%\tabularnewline
\hline 
IBM Brisbane & 14\% & - & -\tabularnewline
\hline 
IBM Fez & - & 5\% & 15\%\tabularnewline
\hline 
IBM Kyiv & 15\% & 9\% & 16\%\tabularnewline
\hline
IONQ Aria1 & 12\% & - & - \tabularnewline
\hline 
\end{tabular}
\par\end{centering}
\caption{\label{tab:achieved_quantum_advantages}Predicted values of the quantum
advantage $A$ for graph domination on 5-, 6-, and 7-site cycles compared
to the values achieved after averaging over $\sim 1000000$
runs of the simulated game using different quantum processors.}
\end{table}
Our explorations on cloud-based quantum hardware show that most of the available quantum advantage can be realised. This is auspicious both for the possibility of realising the advantage in real-life scenarios and using quantum processors to simulate complex graph-domination strategies beyond the reach of classical computers (i.e., those involving large numbers of entangled qubits). Although the limited gate fidelities and finite qubit coherence times of the NISQ devices used here are sufficient to demonstrate the quantum advantage conceptually on small graphs, experimental errors will become an important limiting factor for practical applications. Moreover, current hardware is not yet field-deployable; quantum networking, i.e., the ability to entangle qubits in different processors, will become an essential capability. To this end, significant progress is underway to create room-temperature optical interfaces between remote trapped ions \cite{Main2025, krutyanskiy_entanglement_2023}, neutral atoms \cite{vanLeent2022}, and color centres in diamond \cite{stolk_metropolitan-scale_2024}, as well as cryogenic links between superconducting qubits \cite{magnard_microwave_2020}. However, the simulation of complex strategies will require significantly more high-quality qubits than any current quantum processor or quantum network can provide.

\section{Discussion and Conclusions}

Graph theoretical frameworks are uniquely suited to analyse interconnected systems in a wide range of fields due to their ability to capture complex relationships. Graph domination problems often arise in the context of resource allocation tasks~\cite{Haynes2023-it,FacilityLocation}, where external factors can be dynamically changing. For instance, collision avoidance problems~\cite{CollisionAvoidanceSearch,CollisionAvoidanceSwarm}, can be re-cast as a graph domination game (the problem we have tackled here is indeed collision avoidance on a cycle). In general, the optimal outcome cannot be achieved if real-time communication between the resources is impossible, restricted, unreliable, or unsafe. This issue can arise in critical infrastructure networks, such as ambulances, military units, and electrical power distribution grids.
Using quantum resources, the communication requirements can be reduced \cite{Cleve1997,Xue2001} and, on graphs, nonlocal correlations appear to create the ‘telepathic’ ability that coordinates movements, improving the average payoff. We have identified concrete strategies that realise this operational quantum advantage at dominating circular graphs, and reproduced them successfully on NISQ hardware.

This operational perspective is connected to a broader line of results on coordinating mobile agents~\cite{brukner2006entanglement,RendezvousBase,tucker_quantum-assisted_2024}, including rendezvous games where a quantum advantage was first identified by Mironowicz~\cite{RendezvousBase}. These belong to the family of cooperative \emph{non-local games} (players receive private inputs and cannot communicate) in which entanglement improves performance. Paradigmatic examples include the magic-square game~\cite{Mermin1990,Peres1990} and the odd-cycle game~\cite{Cleve2004}, both demonstrated experimentally~\cite{Xu2022,Drmota2025}. By contrast, \emph{quantum strategic-form games} consider complete-information, simultaneous-move settings~\cite{meyer2000quantum}, with the quantum Prisoner’s Dilemma providing a canonical example featuring a tension between cooperative and selfish incentives~\cite{Eisert1999}. Some games are cooperative in objective yet strategic-form in structure—such as the Kolkata Restaurant Problem~\cite{QGTMultiAgent}—and thus bridge these two perspectives.

Beyond foundations, these distinctions matter for applications. In networked settings, quantum strategies can often be \emph{embedded} into existing classical protocols to yield measurable advantages, particularly when communication is constrained. Entanglement distribution is a key network resource~\cite{ExperimentalTeleportation,QuantumTeleportation1,QGTQuantumNetworks}; the non-local, cooperative games studied here provide primitives for leveraging that resource under minimal signalling. More broadly, the Operational Research literature~\cite{OverviewOR} offers many real-world problems amenable to game-theoretic (including Bayesian) formulations, and several domains—collision avoidance~\cite{CollisionAvoidanceSearch,CollisionAvoidanceSwarm,CollisionAvoidance}, facility location~\cite{FacilityLocation}, and related coverage tasks~\cite{Haynes2023-it}—map naturally to domination on graphs. The techniques presented here invite extensions to non-circular graphs (including mixed degree) and to higher-dimensional entangled resources, suggesting substantial untapped avenues for \emph{operational} quantum advantage.

\section*{Acknowledgements}
CW would like to thank Joshua Tucker and Elizabeth Chipperfield 
for useful discussions. CW acknowledges a studentship awarded by the Engineering and Physical Sciences Research Council (EPSRC) (Grant Number: EP/W52461X/1). This work was supported by the National Quantum Computing Centre through its Quantum Computing Access Program (project number ACA0014).

\newpage
\bibliographystyle{unsrt}
\bibliography{biblio}

@article{tucker_quantum-assisted_2024,
	title = {Quantum-assisted {Rendezvous} on {Graphs}: {Explicit} {Algorithms} and {Quantum} {Computer} {Simulations}},
	issn = {1367-2630},
	shorttitle = {Quantum-assisted {Rendezvous} on {Graphs}},
	url = {http://iopscience.iop.org/article/10.1088/1367-2630/ad78f8},
	doi = {10.1088/1367-2630/ad78f8},
	language = {en},
	urldate = {2024-09-23},
	journal = {New Journal of Physics},
	author = {Tucker, Joshua T. and Strange, Paul and Mironowicz, Piotr and Quintanilla, Jorge},
	year = {2024},
}

@article{PiotrBasePaper,
  title = {Quantum strategies for rendezvous and domination tasks on graphs with mobile agents},
  author = {Viola, Giuseppe and Mironowicz, Piotr},
  journal = {Phys. Rev. A},
  volume = {109},
  issue = {4},
  pages = {042201},
  numpages = {15},
  year = {2024},
  month = {Apr},
  publisher = {American Physical Society},
  doi = {10.1103/PhysRevA.109.042201},
  url = {https://link.aps.org/doi/10.1103/PhysRevA.109.042201}
}

@article{QuantumTeleportation1,
  title = {Teleporting an unknown quantum state via dual classical and Einstein-Podolsky-Rosen channels},
  author = {Bennett, Charles H. and Brassard, Gilles and Cr\'epeau, Claude and Jozsa, Richard and Peres, Asher and Wootters, William K.},
  journal = {Phys. Rev. Lett.},
  volume = {70},
  issue = {13},
  pages = {1895--1899},
  numpages = {0},
  year = {1993},
  month = {Mar},
  publisher = {American Physical Society},
  doi = {10.1103/PhysRevLett.70.1895},
  url = {https://link.aps.org/doi/10.1103/PhysRevLett.70.1895}
}

@Article{ExperimentalTeleportation,
author={Bouwmeester, Dik and Pan, Jian-Wei and Mattle, Klaus
and Eibl, Manfred and Weinfurter, Harald and Zeilinger, Anton},
title={Experimental quantum teleportation},
journal={Nature},
year={1997},
month={Dec},
day={01},
volume={390},
number={6660},
pages={575-579},
issn={1476-4687},
doi={10.1038/37539},
url={https://doi.org/10.1038/37539}
}

@INPROCEEDINGS{ShorAlgorithm,
  author={Shor, P.W.},
  booktitle={Proceedings 35th Annual Symposium on Foundations of Computer Science}, 
  title={Algorithms for quantum computation: discrete logarithms and factoring}, 
  year={1994},
  volume={},
  number={},
  pages={124-134},
  doi={10.1109/SFCS.1994.365700}
}

@inproceedings{GroverAlgoritm,
author = {Grover, Lov K.},
title = {A fast quantum mechanical algorithm for database search},
year = {1996},
isbn = {0897917855},
publisher = {Association for Computing Machinery},
address = {New York, NY, USA},
url = {https://doi.org/10.1145/237814.237866},
doi = {10.1145/237814.237866},
booktitle = {Proceedings of the Twenty-Eighth Annual ACM Symposium on Theory of Computing},
pages = {212–219},
numpages = {8},
location = {Philadelphia, Pennsylvania, USA},
series = {STOC '96}
}

@article{RendezvousBase,
doi = {10.1088/1367-2630/acb22d},
url = {https://dx.doi.org/10.1088/1367-2630/acb22d},
year = {2023},
month = {jan},
publisher = {IOP Publishing},
volume = {25},
number = {1},
pages = {013023},
author = {Mironowicz, P},
title = {Entangled rendezvous: a possible application of Bell non-locality for mobile agents on networks},
journal = {New Journal of Physics}
}

@article{OverviewOR,
author = {Fotios Petropoulos and Gilbert Laporte and Emel Aktas and Sibel A. Alumur and Claudia Archetti and Hayriye Ayhan and Maria Battarra et al.},
title = {Operational Research: methods and applications},
journal = {Journal of the Operational Research Society},
volume = {75},
number = {3},
pages = {423--617},
year = {2024},
publisher = {Taylor \& Francis},
doi = {10.1080/01605682.2023.2253852},
URL = { 
        https://doi.org/10.1080/01605682.2023.2253852
},
eprint = { 
        https://doi.org/10.1080/01605682.2023.2253852
}
}

@book{vedral2006introduction,
  title={Introduction to Quantum Information Science},
  author={Vedral, V.},
  isbn={9780199215706},
  lccn={2007270101},
  series={Introduction to Quantum Information Science},
  url={https://books.google.co.uk/books?id=HNkTDAAAQBAJ},
  year={2006},
  publisher={OUP Oxford}
}

@book{QCIntro,
  title={Quantum Computing: A Gentle Introduction},
  author={Rieffel, E.G. and Polak, W.H.},
  isbn={9780262526678},
  lccn={2010022682},
  series={Scientific and Engineering Computation},
  url={https://books.google.co.uk/books?id=jL34DwAAQBAJ},
  year={2014},
  publisher={MIT Press}
}

@article{meyer2000quantum,
  title={Quantum games and quantum algorithms},
  author={Meyer, David A},
  journal={arXiv preprint quant-ph/0004092},
  year={2000}
}

@misc{QGTQuantumNetworks,
      title={Quantum Game Theory meets Quantum Networks}, 
      author={Indrakshi Dey and Nicola Marchetti and Marcello Caleffi and Angela Sara Cacciapuoti},
      year={2024},
      eprint={2306.08928},
      archivePrefix={arXiv},
      primaryClass={cs.NI},
      url={https://arxiv.org/abs/2306.08928}, 
}

@Inbook{QGTMultiAgent,
author="Sharif, Puya
and Heydari, Hoshang",
editor="Abergel, Fr{\'e}d{\'e}ric
and Chakrabarti, Bikas K.
and Chakraborti, Anirban
and Ghosh, Asim",
title="An Introduction to Multi-player, Multi-choice Quantum Games: Quantum Minority Games {\&} Kolkata Restaurant Problems",
bookTitle="Econophysics of Systemic Risk and Network Dynamics",
year="2013",
publisher="Springer Milan",
address="Milano",
pages="217--236",
isbn="978-88-470-2553-0",
doi="10.1007/978-88-470-2553-0_14",
url="https://doi.org/10.1007/978-88-470-2553-0_14"
}

@ARTICLE{CollisionAvoidance,
  author={Rezaee, Mohammad Reza and Hamid, Nor Asilah Wati Abdul and Hussin, Masnida and Zukarnain, Zuriati Ahmad},
  journal={IEEE Transactions on Intelligent Transportation Systems}, 
  title={Comprehensive Review of Drones Collision Avoidance Schemes: Challenges and Open Issues}, 
  year={2024},
  volume={25},
  number={7},
  pages={6397-6426},
  doi={10.1109/TITS.2024.3375893}}

@book{HaynesHHS1998,
  author    = {Teresa W. Haynes and Stephen T. Hedetniemi and Peter J. Slater},
  title     = {Fundamentals of Domination in Graphs},
  year      = {1998},
  publisher = {Marcel Dekker, Inc.},
  address   = {New York - Basel},
  series    = {Pure and Applied Mathematics - A Series of Monographs and Textbooks}
}

@article{brukner2006entanglement,
  title={Entanglement-assisted orientation in space},
  author={Brukner, {\v{C}}aslav and PAUNKOVI{\'C}, NIKOLA and Rudolph, Terry and Vedral, Vlatko},
  journal={International Journal of Quantum Information},
  volume={4},
  number={02},
  pages={365--370},
  year={2006},
  publisher={World Scientific}
}

@article{Pirandola2020Dec,
	author = {Pirandola, S. and Pirandola, S. and Andersen, U. L. and Banchi, L. and Berta, M. and Bunandar, D. and Colbeck, R. and Englund, D. and Gehring, T. and Lupo, C. and Ottaviani, C. and Pereira, J. L. and Razavi, M. and Shaari, J. Shamsul and Shaari, J. Shamsul and Tomamichel, M. and Tomamichel, M. and Usenko, V. C. and Vallone, G. and Villoresi, P. and Wallden, P.},
	title = {{Advances in quantum cryptography}},
	journal = {Adv. Opt. Photonics},
	volume = {12},
	number = {4},
	pages = {1012--1236},
	year = {2020},
	month = dec,
	issn = {1943-8206},
	publisher = {Optica Publishing Group},
	doi = {10.1364/AOP.361502}
}

@incollection{Klostermeyer2020Oct,
	author = {Klostermeyer, William F. and Mynhardt, C. M.},
	title = {{Eternal and Secure Domination in Graphs}},
	booktitle = {{Topics in Domination in Graphs}},
	journal = {SpringerLink},
	pages = {445--478},
	year = {2020},
	month = oct,
	isbn = {978-3-030-51117-3},
	publisher = {Springer},
	address = {Cham, Switzerland},
	doi = {10.1007/978-3-030-51117-3_13}
}

@article{GoddardHedetniemi2005,
  author  = {Wayne Goddard and Sandra M. Hedetniemi and Stephen T. Hedetniemi},
  title   = {Eternal Security in Graphs},
  journal = {Journal of Combinatorial Mathematics and Combinatorial Computing},
  volume  = {52},
  pages   = {169--180},
  year    = {2005}
}

@inproceedings{chand2023run,
  title={Run for cover: dominating set via mobile agents},
  author={Chand, Prabhat Kumar and Molla, Anisur Rahaman and Sivasubramaniam, Sumathi},
  booktitle={International Symposium on Algorithmics of Wireless Networks},
  pages={133--150},
  year={2023},
  organization={Springer}
}

@misc{Scipy,
  title = {{Scipy minimize} Documentation},
  howpublished = {\url{https://docs.scipy.org/doc/scipy/reference/generated/scipy.optimize.minimize.html}},
  note = {Accessed: 22/09/2025}
}

@BOOK{Haynes2023-it,
  title     = "Domination in graphs: Core concepts",
  author    = "Haynes, Teresa W and Hedetniemi, Stephen T and Henning, Michael
               A",
  publisher = "Springer International Publishing",
  series    = "Springer monographs in mathematics",
  year      =  2023,
  address   = "Cham",
  copyright = "https://www.springernature.com/gp/researchers/text-and-data-mining",
  language  = "en"
}

@INPROCEEDINGS{FacilityLocation,
  author={Sabarish, B. A. and Kailassh, B and Baktha, Kirthana and Janaki, Y},
  booktitle={2017 International Conference on Communication and Signal Processing (ICCSP)}, 
  title={Recommendations of location for facilities using domination set theory}, 
  year={2017},
  volume={},
  number={},
  pages={1540-1544},
  doi={10.1109/ICCSP.2017.8286646}}

@misc{CollisionAvoidanceSearch,
	title = {Graph search algorithms in pathfinding and collision avoidance},
	url = {https://etn-sas.eu/2022/06/14/graph-search-algorithms-in-pathfinding-and-collision-avoidance/},
	language = {nl-BE},
	urldate = {2025-09-30},
	author={Tianlei Miao},
    note={Accessed: 30/09/2025}
}

@INPROCEEDINGS{CollisionAvoidanceSwarm,
  author={Durdu, Akif and Tuncer, Mesut and Yıldız, Berat},
  booktitle={2023 8th International Symposium on Electrical and Electronics Engineering (ISEEE)}, 
  title={Graph-Based Collision Avoidance Algorithm Among Swarm Agents}, 
  year={2023},
  volume={},
  number={},
  pages={19-24},
  doi={10.1109/ISEEE58596.2023.10310351}}

@article{Drmota2025,
	author = {Drmota, P. and Main, D. and Ainley, E. M. and Agrawal, A. and Araneda, G. and Nadlinger, D. P. and Nichol, B. C. and Srinivas, R. and Cabello, A. and Lucas, D. M.},
	title = {{Experimental Quantum Advantage in the Odd-Cycle Game}},
	journal = {Phys. Rev. Lett.},
	volume = {134},
	number = {7},
	pages = {070201},
	year = {2025},
	month = feb,
	publisher = {American Physical Society},
	doi = {10.1103/PhysRevLett.134.070201}
}

@article{Mermin1990,
	author = {Mermin, N. David},
	title = {{Quantum mysteries revisited}},
	journal = {Am. J. Phys.},
	volume = {58},
	number = {8},
	pages = {731--734},
	year = {1990},
	month = aug,
	issn = {0002-9505},
	publisher = {AIP Publishing},
	doi = {10.1119/1.16503}
}

@misc{Peres1990,
	title = {{Incompatible results of quantum measurements}},
	journal = {Phys. Lett. A},
	volume = {151},
	number = {3-4},
	pages = {107--108},
	year = {1990},
	month = dec,
	issn = {0375-9601},
	publisher = {North-Holland},
	note = {[Online; accessed 27. Oct. 2025]},
	doi = {10.1016/0375-9601(90)90172-K}
}

@incollection{Cleve2004,
	author = {Cleve, R. and Hoyer, P. and Toner, B. and Watrous, J.},
	title = {{Consequences and limits of nonlocal strategies}},
	booktitle = {{Proceedings. 19th IEEE Annual Conference on Computational Complexity, 2004.}},
	pages = {24},
	isbn = {978-0-7695-2120},
	publisher = {IEEE},
	doi = {10.1109/CCC.2004.1313847}
}

@article{Xu2022,
	author = {Xu, Jia-Min and Zhen, Yi-Zheng and Yang, Yu-Xiang and Cheng, Zi-Mo and Ren, Zhi-Cheng and Chen, Kai and Wang, Xi-Lin and Wang, Hui-Tian},
	title = {{Experimental Demonstration of Quantum Pseudotelepathy}},
	journal = {Phys. Rev. Lett.},
	volume = {129},
	number = {5},
	pages = {050402},
	year = {2022},
	month = jul,
	publisher = {American Physical Society},
	doi = {10.1103/PhysRevLett.129.050402}
}

@article{Eisert1999,
	author = {Eisert, Jens and Wilkens, Martin and Lewenstein, Maciej},
	title = {{Quantum Games and Quantum Strategies}},
	journal = {Phys. Rev. Lett.},
	volume = {83},
	number = {15},
	pages = {3077--3080},
	year = {1999},
	month = oct,
	publisher = {American Physical Society},
	doi = {10.1103/PhysRevLett.83.3077}
}

@article{Cleve1997,
	author = {Cleve, Richard and Buhrman, Harry},
	title = {{Substituting quantum entanglement for communication}},
	journal = {Phys. Rev. A},
	volume = {56},
	number = {2},
	pages = {1201--1204},
	year = {1997},
	month = aug,
	publisher = {American Physical Society},
	doi = {10.1103/PhysRevA.56.1201}
}

@article{Xue2001,
	author = {Xue, Peng and Huang, Yun-Feng and Zhang, Yong-Sheng and Li, Chuan-Feng and Guo, Guang-Can},
	title = {{Reducing the communication complexity with quantum entanglement}},
	journal = {Phys. Rev. A},
	volume = {64},
	number = {3},
	pages = {032304},
	year = {2001},
	month = aug,
	publisher = {American Physical Society},
	doi = {10.1103/PhysRevA.64.032304}
}

@online{IBMQuantum_Kyiv,
  author       = {{IBM} Quantum},
  title        = {{IBM} Quantum Processor: ibm\_kyiv ({Eagle} r3 Architecture)},
  year         = {2025},
  howpublished = {\url{https://quantum-computing.ibm.com/}},
  note         = {Accessed via the {IBM} Quantum Platform. Processor type: Eagle r3.}
}

@misc{IBMQuantum_Marrakesh,
  author       = {{IBM} Quantum},
  title        = {{IBM} Quantum Processor: ibm\_marrakesh ({Heron} r2 Architecture)},
  howpublished = {\url{https://quantum-computing.ibm.com/}},
  year         = {2025},
  note         = {Accessed via the {IBM} Quantum Platform. Processor type: Heron r1.}
}

@misc{IBMQuantum_Brisbane,
  author       = {{IBM} Quantum},
  title        = {{IBM} Quantum Processor: ibm\_brisbane ({Eagle} r3 Architecture)},
  howpublished = {\url{https://quantum-computing.ibm.com/}},
  year         = {2025},
  note         = {Accessed via the {IBM} Quantum Platform. Processor type: Eagle r3.}
}

@misc{IBMQuantum_Fez,
  author       = {{IBM} Quantum},
  title        = {{IBM} Quantum Processor: ibm\_fez ({Heron} r2 Architecture)},
  howpublished = {\url{https://quantum-computing.ibm.com/}},
  year         = {2025},
  note         = {Accessed via the {IBM} Quantum Platform. Processor type: Heron r1.}
}

@misc{IonQ_Aria1,
  author       = {{IONQ}},
  title        = {{IONQ} Aria1 Quantum Processor},
  howpublished = {\url{https://azure.microsoft.com/en-us/products/quantum-computing/}},
  year         = {2025},
  note         = {Accessed via Microsoft Azure Quantum platform. Processor type: IonQ Aria1.}
}

@article{magnard_microwave_2020,
  title = {Microwave Quantum Link between Superconducting Circuits Housed in Spatially Separated Cryogenic Systems},
  author = {Magnard, P. and Storz, S. and Kurpiers, P. and Sch\"ar, J. and Marxer, F. and L\"utolf, J. and Walter, T. and Besse, J.-C. and Gabureac, M. and Reuer, K. and Akin, A. and Royer, B. and Blais, A. and Wallraff, A.},
  journal = {Phys. Rev. Lett.},
  volume = {125},
  issue = {26},
  pages = {260502},
  numpages = {7},
  year = {2020},
  month = {Dec},
  publisher = {American Physical Society},
  doi = {10.1103/PhysRevLett.125.260502},
  url = {https://link.aps.org/doi/10.1103/PhysRevLett.125.260502}
}

@Article{Main2025,
author={Main, D.
and Drmota, P.
and Nadlinger, D. P.
and Ainley, E. M.
and Agrawal, A.
and Nichol, B. C.
and Srinivas, R.
and Araneda, G.
and Lucas, D. M.},
title={Distributed quantum computing across an optical network link},
journal={Nature},
year={2025},
month={Feb},
day={01},
volume={638},
number={8050},
pages={383-388},
issn={1476-4687},
doi={10.1038/s41586-024-08404-x},
url={https://doi.org/10.1038/s41586-024-08404-x}
}

@article{krutyanskiy_entanglement_2023,
  title = {Entanglement of Trapped-Ion Qubits Separated by 230 Meters},
  author = {Krutyanskiy, V. and Galli, M. and Krcmarsky, V. and Baier, S. and Fioretto, D. A. and Pu, Y. and Mazloom, A. and Sekatski, P. and Canteri, M. and Teller, M. and Schupp, J. and Bate, J. and Meraner, M. and Sangouard, N. and Lanyon, B. P. and Northup, T. E.},
  journal = {Phys. Rev. Lett.},
  volume = {130},
  issue = {5},
  pages = {050803},
  numpages = {7},
  year = {2023},
  month = {Feb},
  publisher = {American Physical Society},
  doi = {10.1103/PhysRevLett.130.050803},
  url = {https://link.aps.org/doi/10.1103/PhysRevLett.130.050803}
}

@article{
stolk_metropolitan-scale_2024,
author = {Arian J. Stolk  and Kian L. van der Enden  and Marie-Christine Slater  and Ingmar te Raa-Derckx  and Pieter Botma  and Joris van Rantwijk  and J. J. Benjamin Biemond  and Ronald A. J. Hagen  and Rodolf W. Herfst  and Wouter D. Koek  and Adrianus J. H. Meskers  and René Vollmer  and Erwin J. van Zwet  and Matthew Markham  and Andrew M. Edmonds  and J. Fabian Geus  and Florian Elsen  and Bernd Jungbluth  and Constantin Haefner  and Christoph Tresp  and Jürgen Stuhler  and Stephan Ritter  and Ronald Hanson },
title = {Metropolitan-scale heralded entanglement of solid-state qubits},
journal = {Science Advances},
volume = {10},
number = {44},
pages = {eadp6442},
year = {2024},
doi = {10.1126/sciadv.adp6442},
URL = {https://www.science.org/doi/abs/10.1126/sciadv.adp6442},
eprint = {https://www.science.org/doi/pdf/10.1126/sciadv.adp6442}}

@Article{vanLeent2022,
author={van Leent, Tim
and Bock, Matthias
and Fertig, Florian
and Garthoff, Robert
and Eppelt, Sebastian
and Zhou, Yiru
and Malik, Pooja
and Seubert, Matthias
and Bauer, Tobias
and Rosenfeld, Wenjamin
and Zhang, Wei
and Becher, Christoph
and Weinfurter, Harald},
title={Entangling single atoms over 33{\thinspace}km telecom fibre},
journal={Nature},
year={2022},
month={Jul},
day={01},
volume={607},
number={7917},
pages={69-73},
issn={1476-4687},
doi={10.1038/s41586-022-04764-4},
url={https://doi.org/10.1038/s41586-022-04764-4}
}
\newpage

\appendix

\section{$C_{10}$ Domination Table}

\label{C10Table}
\begin{table}[h!]
\centering
\begin{tabular}{cl|ll|ll|ll|ll|ll|ll|ll|ll|ll|ll|}
\cline{3-22}
\multicolumn{2}{l|}{\multirow{2}{*}{}} & \multicolumn{2}{c|}{1} & \multicolumn{2}{c|}{2} & \multicolumn{2}{c|}{3} & \multicolumn{2}{c|}{4} & \multicolumn{2}{c|}{5} & \multicolumn{2}{c|}{6} & \multicolumn{2}{c|}{7} & \multicolumn{2}{c|}{8} & \multicolumn{2}{c|}{9} & \multicolumn{2}{c|}{10} \\ \cline{3-22} 
\multicolumn{2}{l|}{} & \multicolumn{1}{c|}{0} & \multicolumn{1}{c|}{1} & \multicolumn{1}{c|}{0} & \multicolumn{1}{c|}{1} & \multicolumn{1}{c|}{0} & \multicolumn{1}{c|}{1} & \multicolumn{1}{c|}{0} & \multicolumn{1}{c|}{1} & \multicolumn{1}{c|}{0} & \multicolumn{1}{c|}{1} & \multicolumn{1}{c|}{0} & \multicolumn{1}{c|}{1} & \multicolumn{1}{c|}{0} & \multicolumn{1}{c|}{1} & \multicolumn{1}{c|}{0} & \multicolumn{1}{c|}{1} & \multicolumn{1}{c|}{0} & \multicolumn{1}{c|}{1} & \multicolumn{1}{c|}{0} & \multicolumn{1}{c|}{1} \\ \hline
\multicolumn{1}{|c|}{\multirow{2}{*}{1}} & 0 & 3 & 5 & 4 & 6 & 5 & 6 & 6 & 6 & 6 & 6 & 6 & 6 & 6 & 5 & 6 & 4 & 5 & 3 & 4 & 4 \\ \cline{2-2}
\multicolumn{1}{|c|}{} & 1 & 5 & 3 & 4 & 4 & 3 & 5 & 4 & 6 & 5 & 6 & 6 & 6 & 6 & 6 & 6 & 6 & 6 & 5 & 6 & 4 \\ \hline
\multicolumn{1}{|c|}{\multirow{2}{*}{2}} & 0 & 4 & 4 & 3 & 5 & 4 & 6 & 5 & 6 & 6 & 6 & 6 & 6 & 6 & 6 & 6 & 5 & 6 & 4 & 5 & 3 \\ \cline{2-2}
\multicolumn{1}{|c|}{} & 1 & 6 & 4 & 5 & 3 & 4 & 4 & 3 & 5 & 4 & 6 & 5 & 6 & 6 & 6 & 6 & 6 & 6 & 6 & 6 & 5 \\ \hline
\multicolumn{1}{|c|}{\multirow{2}{*}{3}} & 0 & 5 & 3 & 4 & 4 & 3 & 5 & 4 & 6 & 5 & 6 & 6 & 6 & 6 & 6 & 6 & 6 & 6 & 5 & 6 & 4 \\ \cline{2-2}
\multicolumn{1}{|c|}{} & 1 & 6 & 5 & 6 & 4 & 5 & 3 & 4 & 4 & 3 & 5 & 4 & 6 & 5 & 6 & 6 & 6 & 6 & 6 & 6 & 6 \\ \hline
\multicolumn{1}{|c|}{\multirow{2}{*}{4}} & 0 & 6 & 4 & 5 & 3 & 4 & 4 & 3 & 5 & 4 & 6 & 5 & 6 & 6 & 6 & 6 & 6 & 6 & 6 & 6 & 5 \\ \cline{2-2}
\multicolumn{1}{|c|}{} & 1 & 6 & 6 & 6 & 5 & 6 & 4 & 5 & 3 & 4 & 4 & 3 & 5 & 4 & 6 & 5 & 6 & 6 & 6 & 6 & 6 \\ \hline
\multicolumn{1}{|c|}{\multirow{2}{*}{5}} & 0 & 6 & 5 & 6 & 4 & 5 & 3 & 4 & 4 & 3 & 5 & 4 & 6 & 5 & 6 & 6 & 6 & 6 & 6 & 6 & 6 \\ \cline{2-2}
\multicolumn{1}{|c|}{} & 1 & 6 & 6 & 6 & 6 & 6 & 5 & 6 & 4 & 5 & 3 & 4 & 4 & 3 & 5 & 4 & 6 & 5 & 6 & 6 & 6 \\ \hline
\multicolumn{1}{|c|}{\multirow{2}{*}{6}} & 0 & 6 & 6 & 6 & 5 & 6 & 4 & 5 & 3 & 4 & 4 & 3 & 5 & 4 & 6 & 5 & 6 & 6 & 6 & 6 & 6 \\ \cline{2-2}
\multicolumn{1}{|c|}{} & 1 & 6 & 6 & 6 & 6 & 6 & 6 & 6 & 5 & 6 & 4 & 5 & 3 & 4 & 4 & 3 & 5 & 4 & 6 & 5 & 6 \\ \hline
\multicolumn{1}{|c|}{\multirow{2}{*}{7}} & 0 & 6 & 6 & 6 & 6 & 6 & 5 & 6 & 4 & 5 & 3 & 4 & 4 & 3 & 5 & 4 & 6 & 5 & 6 & 6 & 6 \\ \cline{2-2}
\multicolumn{1}{|c|}{} & 1 & 5 & 6 & 6 & 6 & 6 & 6 & 6 & 6 & 6 & 5 & 6 & 4 & 5 & 3 & 4 & 4 & 3 & 5 & 4 & 6 \\ \hline
\multicolumn{1}{|c|}{\multirow{2}{*}{8}} & 0 & 6 & 6 & 6 & 6 & 6 & 6 & 6 & 5 & 6 & 4 & 5 & 3 & 4 & 4 & 3 & 5 & 4 & 6 & 5 & 6 \\ \cline{2-2}
\multicolumn{1}{|c|}{} & 1 & 4 & 6 & 5 & 6 & 6 & 6 & 6 & 6 & 6 & 6 & 6 & 5 & 6 & 4 & 5 & 3 & 4 & 4 & 3 & 5 \\ \hline
\multicolumn{1}{|c|}{\multirow{2}{*}{9}} & 0 & 5 & 6 & 6 & 6 & 6 & 6 & 6 & 6 & 6 & 5 & 6 & 4 & 5 & 3 & 4 & 4 & 3 & 5 & 4 & 6 \\ \cline{2-2}
\multicolumn{1}{|c|}{} & 1 & 3 & 5 & 4 & 6 & 5 & 6 & 6 & 6 & 6 & 6 & 6 & 6 & 6 & 5 & 6 & 4 & 5 & 3 & 4 & 4 \\ \hline
\multicolumn{1}{|c|}{\multirow{2}{*}{10}} & 0 & 4 & 6 & 5 & 6 & 6 & 6 & 6 & 6 & 6 & 6 & 6 & 5 & 6 & 4 & 5 & 3 & 4 & 4 & 3 & 5 \\ \cline{2-2}
\multicolumn{1}{|c|}{} & 1 & 4 & 4 & 3 & 5 & 4 & 6 & 5 & 6 & 6 & 6 & 6 & 6 & 6 & 6 & 6 & 5 & 6 & 4 & 5 & 3 \\ \hline
\end{tabular}
\caption{The domination table for the cycle 10 graph where players independently decide whether to move clockwise or anti-clockwise by flipping a coin (or examining a qubit). The number at the top of each column represents Alice's site, and the number at the front of each row represents Bob's site}
\label{C10QuantTable}
\end{table}

\end{document}